\documentclass[journal]{IEEEtran}
\usepackage[noadjust]{cite}
\usepackage{amsmath,amsthm,graphicx,cite,booktabs,amssymb,gensymb, color, xcolor, fancyhdr,float,perpage,tikz, colortbl}
\usepackage[linesnumbered,vlined,ruled]{algorithm2e}
\usepackage{algorithmic,array,enumerate,rotating}
\usepackage{subfigure,epsfig,dblfloatfix}
\usepackage{multirow}
\usepackage[affil-it]{authblk}
\usepackage{hyperref}
\usetikzlibrary{matrix}
\hypersetup{
	colorlinks   = true,
	linkcolor    = blue,
	citecolor    = blue
}

\makeatletter

\renewcommand{\@algocf@capt@plain}{above}

\definecolor{LightGray}{gray}{0.85}
\definecolor{Gray}{gray}{0.65}

\makeatother
\def\bsx{{\boldsymbol{x}}}
\def\bsy{{\boldsymbol{y}}}
\def\bsH{{\boldsymbol{H}}}
\def\bsJ{{\boldsymbol{J}}}
\def\bsI{{\boldsymbol{I}}}

\def\bsV{{\boldsymbol{V}}}
\def\bsA{{\boldsymbol{A}}}
\def\bsB{{\boldsymbol{B}}}
\def\bsc{{\boldsymbol{c}}}
\def\bsR{{\boldsymbol{R}}}
\def\bpsi{{\boldsymbol{\psi}}}

\title{Image Denoising Inspired by Quantum Many-Body physics}
\author{Sayantan~Dutta$^{1,2}$,~Adrian~Basarab$^{1}$,~Bertrand Georgeot$^{2}$,~and~Denis~Kouam\'e$^{1}$}

\affil{
   {\em $^{1}$Institut de Recherche en Informatique de Toulouse, UMR CNRS 5505, Universit\'e de Toulouse, France} \\
  {\em  $^{2}$Laboratoire de Physique Th\'eorique, Universit\'e de Toulouse, CNRS, UPS, France} \\
  {\em  Emails: sayantan.dutta@irit.fr, adrian.basarab@irit.fr, georgeot@irsamc.ups-tlse.fr, denis.kouame@irit.fr}
 }
 
\begin{document}

\pagenumbering{gobble}

\maketitle

\begin{abstract}
Decomposing an image through Fourier, DCT or wavelet transforms is still a common approach in digital image processing, in number of applications such as denoising. In this context, data-driven dictionaries and in particular exploiting the redundancy withing patches extracted from one or several images allowed important improvements. This paper proposes an original idea of constructing such an image-dependent basis inspired by the principles of quantum many-body physics. The similarity between two image patches is introduced in the formalism through a term akin to interaction terms in quantum mechanics. The main contribution of the paper is thus to introduce this original way of exploiting quantum many-body ideas in image processing, which opens interesting perspectives in image denoising. The potential of the proposed adaptive decomposition is illustrated through image denoising in presence of additive white Gaussian noise, but the method can be used for other types of noise such as image-dependent noise as well. Finally, the results show that our method achieves comparable or slightly better results than existing approaches.
\end{abstract}


\vspace{-4mm}
\section{Introduction}
\label{sec:intro}

Transforming a noisy image into a sparse representation using Fourier, DCT or wavelet basis vectors still remains a major step for image restoration problems due to their ability to preserve most of the image energy into few coefficients. From these coefficients the clean image can be efficiently estimated \cite{donoho1994ideal, donoho1995wavelet, starck2002curvelet}.
During the past two decades, the redundancy between patches extracted from one or several images has been shown to be a key aspect for number of denoising techniques like non-local means (NLM) \cite{buades2005review}, dictionary learning \cite{Aharon2006An,Elad2006image} or block-matching and 3D filtering (BM3D) \cite{Dabov2007image}. These data-driven methods are designed to use the information from neighbouring image-patches to preserve the local structures of the target image.

In this paper, we propose a novel idea of image representation using the concepts of quantum many-body interaction \cite{mahan2013many}. Recently, several attempts of implementing quantum principles in imaging application have been initiated, particularly for image segmentation \cite{Aytekin2013Quantum, youssry2015quantum, youssry2019continuous} or denoising \cite{kaisserli2014image, kaisserli2015novel, dutta2020quantum, Smith2018Adaptive}. Alternatively, designing image processing schemes for quantum computers has been also largely explored (see, e.g., \cite{iliyasu2013towards, zhang2013neqr}), but is out of the scope of this work. Most of these quantum image processing methods use the theory of quantum mechanics for a single particle system to provide an adaptive basis. In this present paper, we generalize such methods using tools inspired by quantum many-body physics, which enable to get a more versatile adaptive basis which takes into account similarities between neighbouring image patches.


The image denoising problem using quantum principles has been already discussed in some of our previous works \cite{dutta2020quantum, Smith2018Adaptive}, based on the single-particle framework. Although this adaptive method has been shown to be very efficient for different types of noise, it still faces some challenges, among which the most important are: i) computational burden limits its application to large scale images, 
ii) the method does not take advantage of the structural properties of the image, which are known to be useful in many algorithms. In this paper, we propose to mitigate these drawbacks by exploiting ideas of quantum many-body physics for constructing an adaptive denoising method.

Interactions in quantum physics correspond to two or more quantum particles present in the system that can influence each other's quantum state. From an image processing perspective, we propose to adapt this theory to extend the idea of interaction between image patches. More precisely, the proposed framework consists in placing a quantum particle in every image-patch, \textit{i.e.,} every image-patch acts like a single particle system, and the whole collection of patches, \textit{i.e.} the image, behaves like a many-body system where interactions describe local similarities in the neighbouring patches. 

The paper is organized as follows. We discuss the concepts of quantum many-body interaction in Section~\ref{sec:qmbi} and develop the idea of many-body interaction from an image processing perspective in Section~\ref{sec:qmbiVSim}. The proposed image denoising algorithm is described in Section~\ref{sec:algorithm}. Numerical experiments are reported in Section~\ref{sec:results} showing the efficiency of the proposed method, before conclusions and perspectives in Section~\ref{sec:conclusion}.

\begin{figure*}[h!]
\centering
\includegraphics[width=1\textwidth]{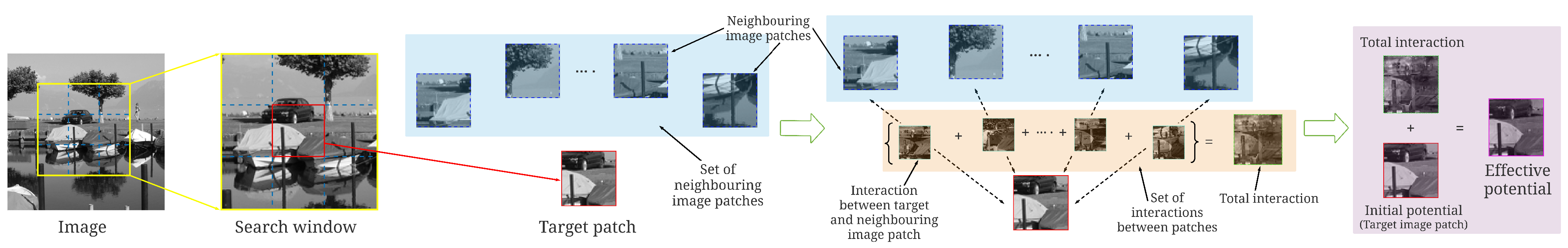}
\vspace{-8mm}
\caption{Many-body interaction from an image processing viewpoint.}
\label{fig:def_inter}
\vspace{-7mm}
\end{figure*}

\vspace{-3mm}
\section{Quantum many-body interaction}
\label{sec:qmbi}

Quantum mechanics describes our world at a fundamental level, and classical mechanics is merely an approximation of quantum theory in a certain limit.  In classical theory, the position of a particle is always determined precisely, whereas in quantum theory only probabilities of presence can be computed.
The basic object in quantum mechanics is the wave function, whose modulus square gives the probability of presence of a particle, and which obeys a wave equation called the Schroedinger equation.
Thus, for a non-relativistic single particle system, the probability of presence of a particle with energy $E$ in a potential $V(y)$ is determined by the wave function $\psi(y)$, where $y$ defines the spatial coordinate. This wave function belongs to the Hilbert space of $L^2$-integrable functions and obeys the stationary Schroedinger equation:
\begin{equation}
 - \frac{\hbar ^2}{2m} \nabla_y ^2 \psi (y) = - V(y)  \psi (y) + E \psi (y),
\label{eq:schroedinger}
\end{equation}
\noindent with $m$ the mass of the quantum particle, $\hbar$ the Planck constant and $\nabla_y$ the gradient operator. In operator notation, \eqref{eq:schroedinger} corresponds to $H \psi (y) = E \psi (y) $ with the Hamiltonian operator $H = -(\hbar ^2/2m) \nabla_y^2 + V(y)$. The function $|\psi (y)|^2$ gives the probability of finding the particle at some point.

For imaging applications, this Hamiltonian operator $H$ has been discretized and implemented as a tool for constructing an adaptive basis as proposed in \cite{dutta2020quantum, Smith2018Adaptive}. This discretized Hamiltonian operator reads as:
\begin{eqnarray}
\bsH[i,j]= \left \{
   \begin{array}{c c l}
      \bsx[i]+ 4 \frac{\hbar ^2}{2m} &  & for \; i=j,\\
       -\frac{\hbar ^2}{2m} & & for \; i = j \pm 1,\\
        -\frac{\hbar ^2}{2m} & & for \; i = j \pm n,\\
      0 & & otherwise.
   \end{array}
   \right.
\label{eq:H}
\end{eqnarray}
where $\bsx \in {\rm I \!R}^{n^2}$ is an image (\textit{i.e.,} $V=\bsx$), and $\bsH [i,j]$ and $\bsx[i]$ represent respectively the $(i,j)$-th component of the Hamiltonian operator and the $i$-th component of the vectorized image $\bsx$ in the lexicographical order. Note that zero-padding is used to handle the boundary conditions \cite{dutta2020quantum}. The corresponding set of eigenvectors of the Hamiltonian operator \eqref{eq:H} represents an adaptive basis on which the image is decomposed prior to denoising \cite{dutta2020quantum, Smith2018Adaptive}. This basis corresponds to the stationary solutions of the Schroedinger equation in a potential given by the pixel values. These basis vectors are oscillatory functions with oscillation frequencies depending on the local value of the potential (\textit{i.e.,} image pixels).

Quantum mechanics can be generalized to more than one particle. Let us denote by $z$ the number of particles in a quantum system. As a consequence, a particle-to-particle interaction takes place inside the system and the Hamiltonian operator for the many-body system becomes \cite{mahan2013many}:
\begin{equation}
H = - \sum_{a=1}^z \dfrac{\hbar ^2}{2m_a} \nabla_{y_a}^2 + \dfrac{1}{2} \sum_{a=1}^z \sum_{b=1, b\neq a}^z V_{ab},
\label{eq:H_mb}
\end{equation}
where $V_{ab}$ is a function of $ y_1,y_2,\cdots,y_z $, the positions of the $z$ particles. The difficulty of solving the many-body problem \eqref{eq:H_mb}  comes from the complex nature of $V_{ab}$, which can be tackled under some approximations from quantum theory.






We propose to use ideas from this theory to incorporate similarities between patches in the quantum formalism producing the adaptive basis for imaging applications. Therefore, we divide the image into patches indexed from $1$ to $z$, and write the Hamiltonian in each patch as

\begin{equation}
H_a = \underbrace{  \vphantom{ \sum_{b=1, b\neq a}^z I_{ab}}
-\dfrac{\hbar ^2}{2m_a} \nabla_{y_a}^2  + V(y_a) }_{H_{0_a}} + \underbrace{ \sum_{b=1, b\neq a}^z I_{ab} }_{H_{I_a}}, ~ a = 1,\cdots,z
\label{eq:H_hfmf}
\end{equation}
where, $I_{ab}$ is the interaction between the $a$-th and $b$-th patches, $H_{0_a}$ is the Hamiltonian in the $a$-th patch as a single particle system (as it appears discretized in \eqref{eq:H}), and $H_{I_a}$ is the total interaction between the $a$-th patch and the other patches in the system. So the effective potential $V_a^{effective}$ inside the $a$-th patch is
\begin{equation}
V_a^{effective} = V(y_a) + \sum_{b=1, b\neq a}^z I_{ab} = V(y_a) + H_{I_a}.
\label{eq:V_eff}
\end{equation}


The problem of finding the adaptive basis transfers thus into solving $z$ systems of equations, as follows:
\begin{equation}
H_a \psi(y_a) = E_a \psi(y_a), ~~~ a = 1,2,\cdots,z.
\label{eq:schro_hfmf}
\end{equation}
which should be discretized in each patch as in  \eqref{eq:H}.

\section{Quantum many-body interaction from image processing viewpoint}
\label{sec:qmbiVSim}

\subsection{General framework}
\label{sec:genfram}

We propose to relate many-body interaction systems to image processing under the following principles:

\begin{itemize}

\item The image (the pixel values) acts as the potential for a quantum system.

\item For a single-particle system, the probability of presence of a quantum particle at some point on the potential, \textit{i.e.} image, is governed by the wave function $\bpsi (y)$.

\item This wave function $\bpsi (y)$ is a solution of \eqref{eq:schroedinger}, while the image acts as the potential $\bsV(y)$.

\item This wave function belongs to the set of oscillatory functions, with local frequencies dependent on the image pixels values. The oscillation frequency is low for higher values of the pixels and vice-versa.

\item An imaginary quantum particle is associated to every small patch extracted from an image. Each of these potential surfaces with a quantum particle behaves as a single-particle system.

\item These single-particle systems are not independent, but interactions occur between them and the other patches inside the whole image, similar to what happens in a quantum many-body system, where each quantum particle interacts with other quantum particles present in the system (see Fig.~\ref{fig:def_inter}).

\item These quantum interactions modify the effective potential of the quantum particle following \eqref{eq:V_eff}. Indeed, the shape of the wave function depends on these interactions.

\end{itemize}

\subsection{Quantum interaction between two image patches}
\label{sec:definter}

In nature, four fundamental interactions exist: gravitational, electromagnetic, strong, and weak interactions. Mathematically, the two interactions which are not short-range correspond to an inverse-square law. Without loss of generality, we propose to extend this idea to image processing, as follows:

\begin{itemize}

\item The interaction between two image patches is inversely proportional to the square of the physical or Euclidean distance between the patches, \textit{i.e.,}
$\bsI_{ab} \propto \frac{1}{D_{ab}^2}$,
where $\bsI_{ab}$ and $D_{ab}$ are respectively the interaction and the Euclidean distance between two patches denoted by $\bsA$ and $\bsB$.

\item The interaction between two image patches is linearly proportional to the absolute value of the pixel-wise difference between the patches. This process is defined pixel-wise, \textit{i.e.,}
$\bsI_{ab}^i \propto |\bsA^i - \bsB^i|$, $i = 1,2,\cdots,P_{dim}$,
where superscript $i$ indicates the $i$-th element and $P_{dim}$ is the number of pixels in every image patch.

\end{itemize}

Hence, in image processing, for an interacting many-patch system the inverse-square law can be defined as
\begin{equation}
\bsI_{ab}^i =  p \frac{|\bsA^i - \bsB^i|}{D_{ab}^2} ,~~ i = 1,2,\cdots,P_{dim},
\label{eq:invsqulaw}
\end{equation}
where $p$ is a proportionality constant, which will ultimately act as a hyperparameter for our problem.

\vspace{-4mm}
\subsection{Interpretation of the inverse-square law for image patches}
\label{sec:interpre}

The proposed inverse-square law for a many-patch interaction model can be interpreted in the following manners: i) if pixel values of two patches are very similar then they are less interactive, ii) if two patches are similar but placed far from each other in the image then they present small-scale interaction. In other words, if neighbouring patches are very different from each other then they exhibit high interactions, but distant patches have always low interaction in spite of their possible dissimilarity. 
The interactions between the target patch and its neighbouring patches manifest themselves in a way such that the effective potential is obtained by adding the initial potential (\textit{i.e.,} the target patch itself) with the total interaction term, thus incorporating the idea of patch similarity in the local neighbourhood. Fig.~\ref{fig:def_inter} depicts a visual representation of the proposed methodology relating patch-wise image processing and quantum many-body interaction.
We note that other laws than the inverse square law can be used, this amounts to modifying the importance of distant patches compared to neighbouring ones in the algorithm.

\vspace{-3mm}
\section{Quantum interactions in image denoising}
\label{sec:algorithm}

The proposed idea of quantum interactive patches can be explored, for example, to address an image denoising problem. In this context, the primary objective is to construct an adaptive basis for each individual patch, which will be further used in the decomposition of that patch. These basis vectors for the $k$-th patch, are the solutions of \eqref{eq:schroedinger} with the effective potential denoted by $\bsV_k^{effective}$ in \eqref{eq:V_eff}. In other words, these basis vectors are the eigenvectors of the Hamiltonian matrix \eqref{eq:H} under the effective potential $\bsV_k^{effective}$, that represents the sum between the current patch and its interactions with its neighbouring patches.

These basis vectors are oscillating functions with: i) oscillation frequency increasing with energy (\textit{i.e.,} eigenvalue in \eqref{eq:schro_hfmf}), and ii) a given basis vector having low local frequencies for high values of the effective potential $\bsV_k^{effective}$ and vice-versa. 
Patch denoising can be achieved by projecting the noisy patch onto a $d$-dimensional subspace corresponding to the solutions of \eqref{eq:schro_hfmf} of lowest energies, and reconstruct the denoised patch using these projection coefficients. Here, $d$ acts as a threshold. In this way, a lack of similarity between pixels leads to a stronger denoising, since for the same value of the energy these regions will have lower frequencies than the ones with more similarity. Finally combining all the denoised patches, similar to standard non-local means algorithms, one can obtain the denoised image. The proposed denoising algorithm is resumed in Algo.~\ref{Algo:QMBI}.

\vspace{-3mm}

\begin{algorithm}[h!]
\begin{footnotesize}
\label{Algo:QMBI}

\KwIn{ $\bsy$ , $P_h$, $W_h$, $d$, $p$, $\frac{\hbar ^2}{2m}$}

 {Divide the noisy image $\bsy$ into $T_{patch}$ small patches of size $(2*P_h + 1)$. \textit{i.e.,} $P_{dim} = (2*P_h + 1)^2$}\\

  \For{ $w = 1 : T_{patch}$}{
  
 {Choose one image patch $\bsJ_w$}\\
 
 {Create a search window of size $(2*W_h + 1)$ centered on $\bsJ_w$}\\
 
 {Collect all $S_{patch}$ image patches inside this search window}\\

     \For{ $l = 1 : S_{patch}$}{
     
     {Calculate Euclidean distance $D_{wl}$ between $\bsJ_w$ and $\bsJ_l$ patches inside the search window}\\
     
     {Calculate interaction $\bsI_{wl}$ between $\bsJ_w$ and $\bsJ_l$ patches inside the search window as $\bsI_{wl}^k = p \dfrac{ | \bsJ_w^k - \bsJ_l^k | }{ D_{wl}^2}, ~~k = 1,\cdots, P_{dim}$}\\
         
     } 
 
 {Calculate total interaction $\bsI^{~total}_w$ between the patch $\bsJ_w$ and patches inside the search window by taking the sum over all $l$ \textit{i.e.,} $\bsI^{{~total}^k}_w = \sum_{l = 1}^{S_{patch}}  \bsI_{wl}^k , ~~ k = 1,\cdots,P_{dim}$}\\
  
 {Effective potential for the $\bsJ_w$ patch is $ \bsV_w^{{~effective}^k} = \bsJ_w^k + \bsI^{{~total}^k}_w , ~~ k = 1,\cdots,P_{dim}$}\\
 
 {Construct the Hamiltonian matrix $\bsH_w$ using the effective potential $ \bsV_w^{~effective}$}\\
 
 {Calculate the eigenvalues and eigenvectors of $\bsH_w$}\\
 
 {Construct adaptive basis $\bsB_w^{adaptive}$ using the eigenvectors $\bpsi_w^k , ~~ k = 1,\cdots,P_{dim}$}\\

     {Project the noisy patch $\bsJ_w$ onto this adaptive basis $\bsB_w^{adaptive}$}\\
     
     {Calculate projection coefficients $\bsc_w$ in the $P_{dim}$-dimensional space. Note that, $P_{dim} > d$}\\
     
     {Redefine the projection coefficients in the $d$-dimensional subspace as $\bsc^{{new}^k}_{w} = \bsc^k_{w} , k = 1,\cdots,d$}\\
     
     {Reconstruct the patch by $\bsR_w = \sum_{k = 1}^d \bsc^{{new}^k}_{w} \bpsi_w^k$}\\
          
 }
 
 {Combine all $T_{patch}$ denoised patches $\bsR_w$, to obtain the full denoised image $\hat{\bsx}$}\\
 
 \KwOut{$\hat{\bsx}$}

\caption{QMPI denoising algorithm}

\DecMargin{1em}
\end{footnotesize}

\end{algorithm}

\vspace{-8mm}

\section{Simulation results}
\label{sec:results}

This section provides the numerical experiments conducted with the proposed adaptive quantum many-patch interaction (QMPI) method for image denoising. We illustrate the efficiency through three standard images (house, lake and Lena) corrupted by additive whithe Gaussian noise (AWGN) corresponding to different levels of signal-to-noise-ratio (SNR) (22, 16, 8, and 2 dB).

\begin{figure*}[h!]
\centering

\subfigure[{\scriptsize Clean image}]{\includegraphics[width=0.16\textwidth]{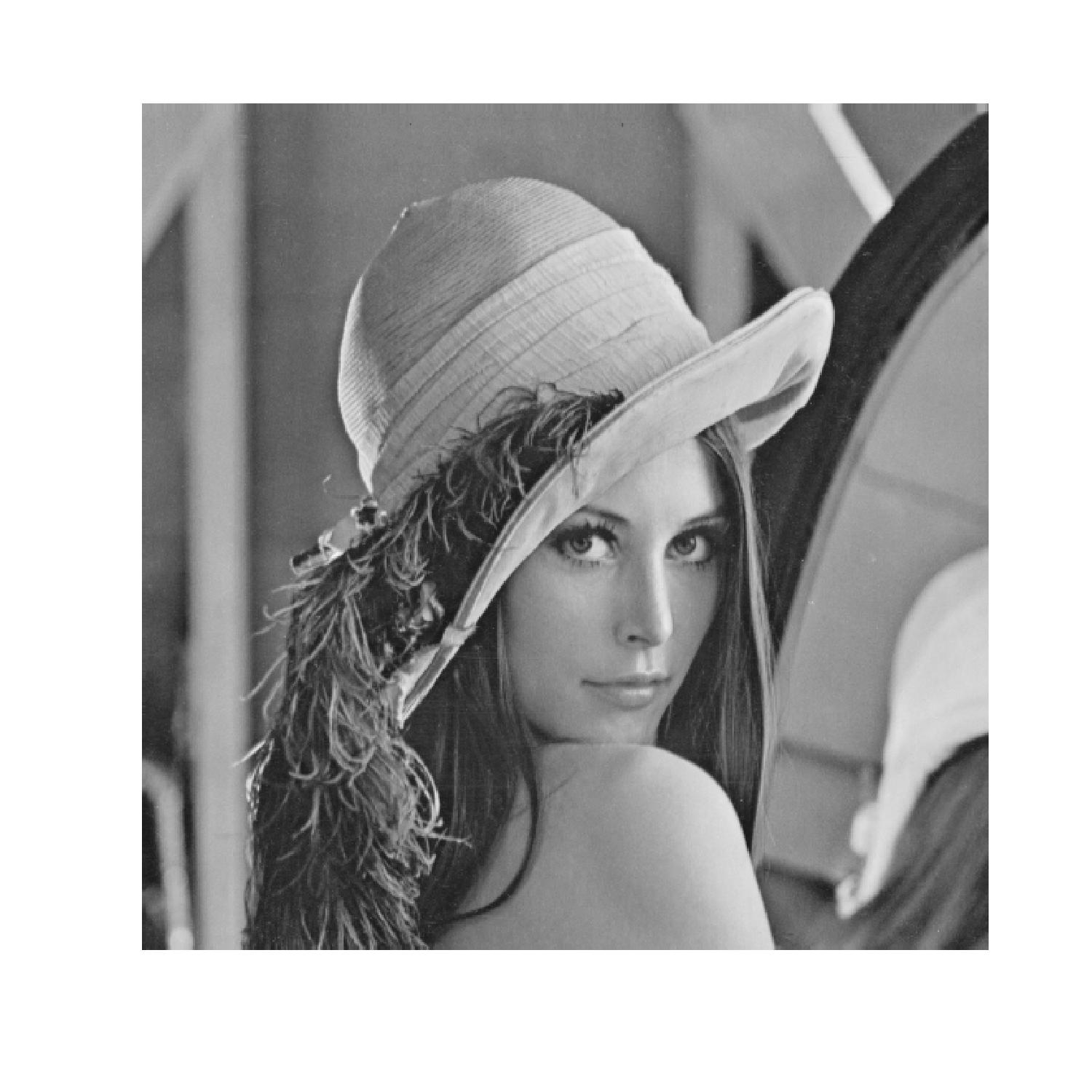}}
\subfigure[{\scriptsize Noisy image(16dB)}]{\includegraphics[width=0.16\textwidth]{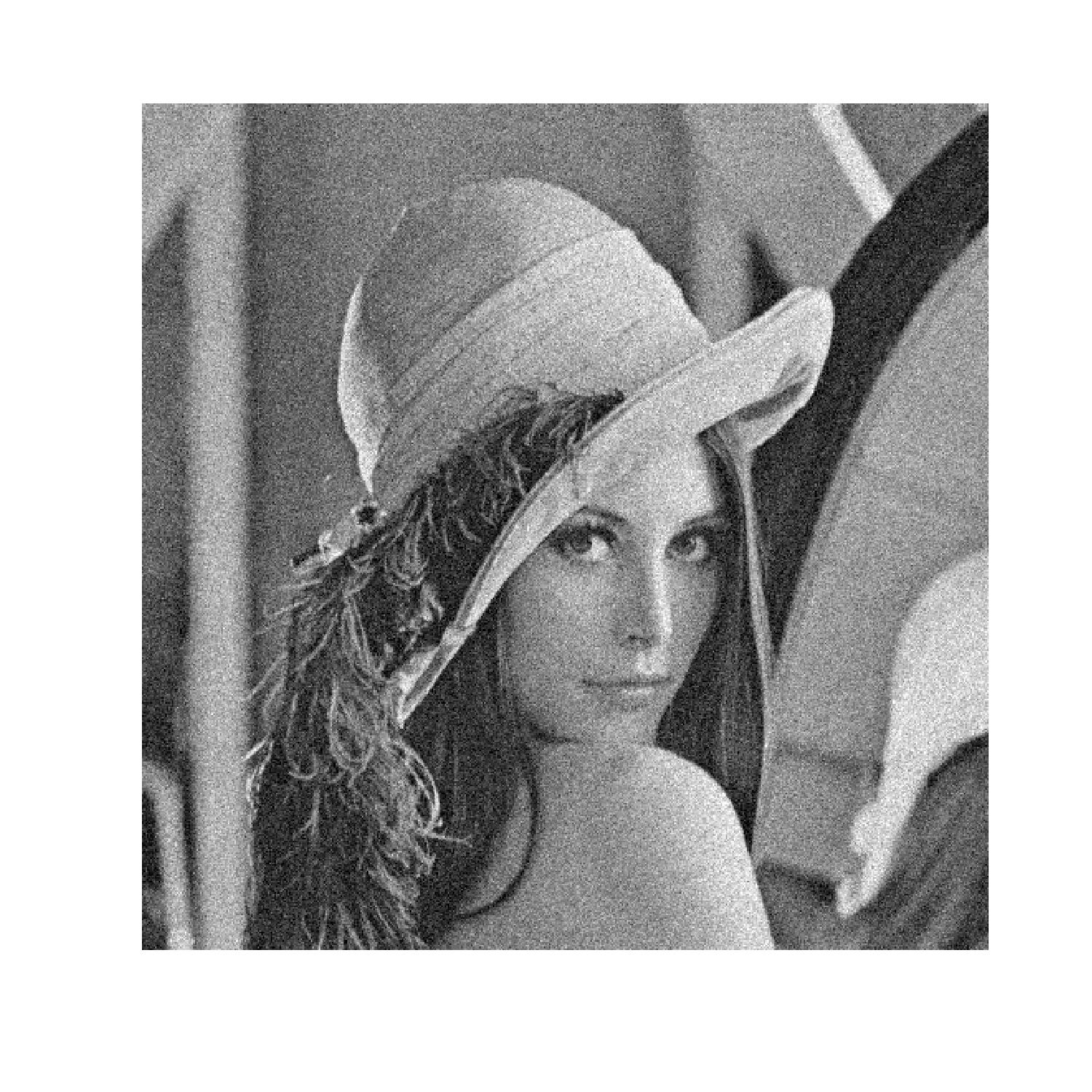}}
\subfigure[{\scriptsize PND(31.32dB/0.828)}]{\includegraphics[width=0.16\textwidth]{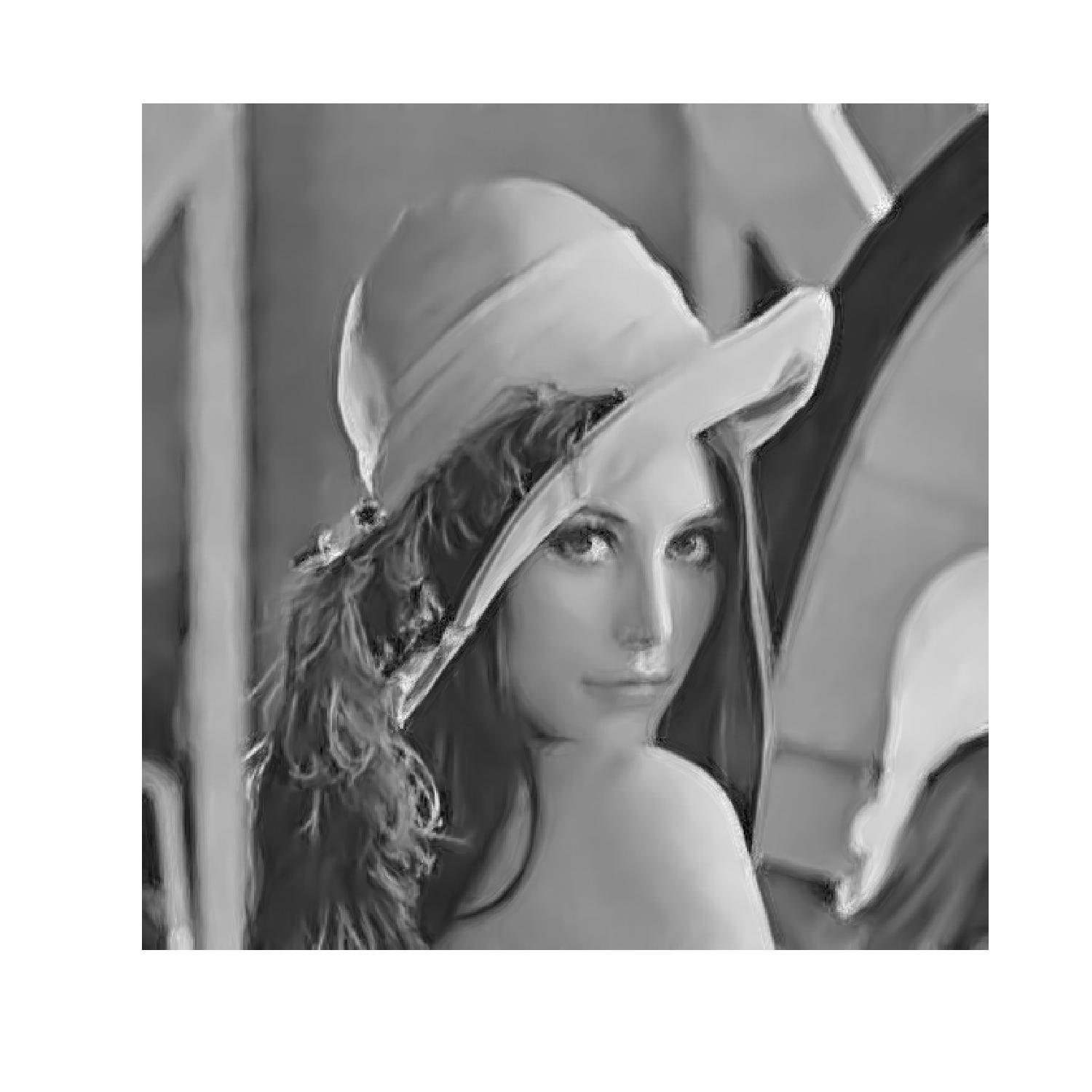}}
\subfigure[{\scriptsize PGPCA(31.81dB/0.815)}]{\includegraphics[width=0.16\textwidth]{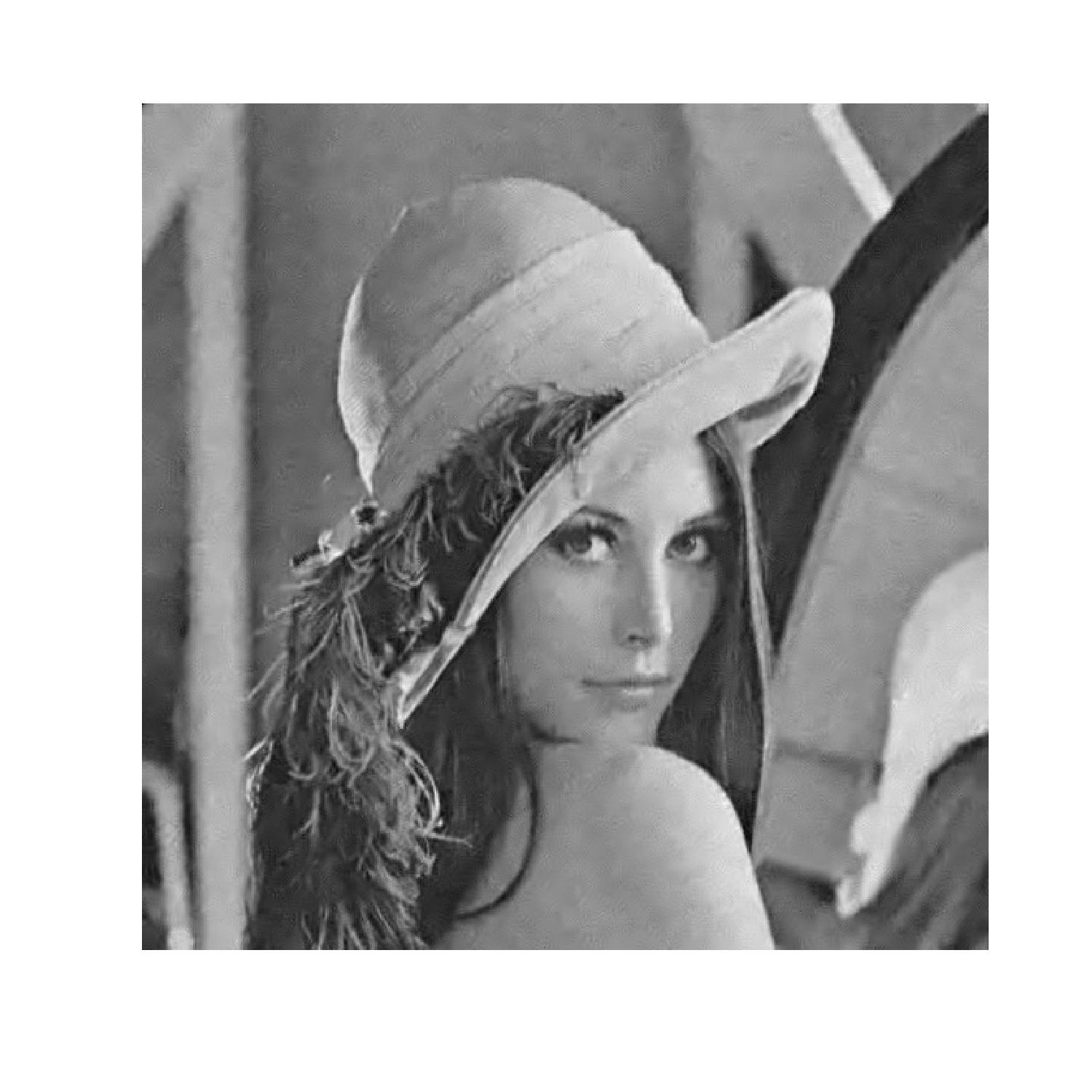}}
\subfigure[{\scriptsize PLPCA(31.89dB/0.806)}]{\includegraphics[width=0.16\textwidth]{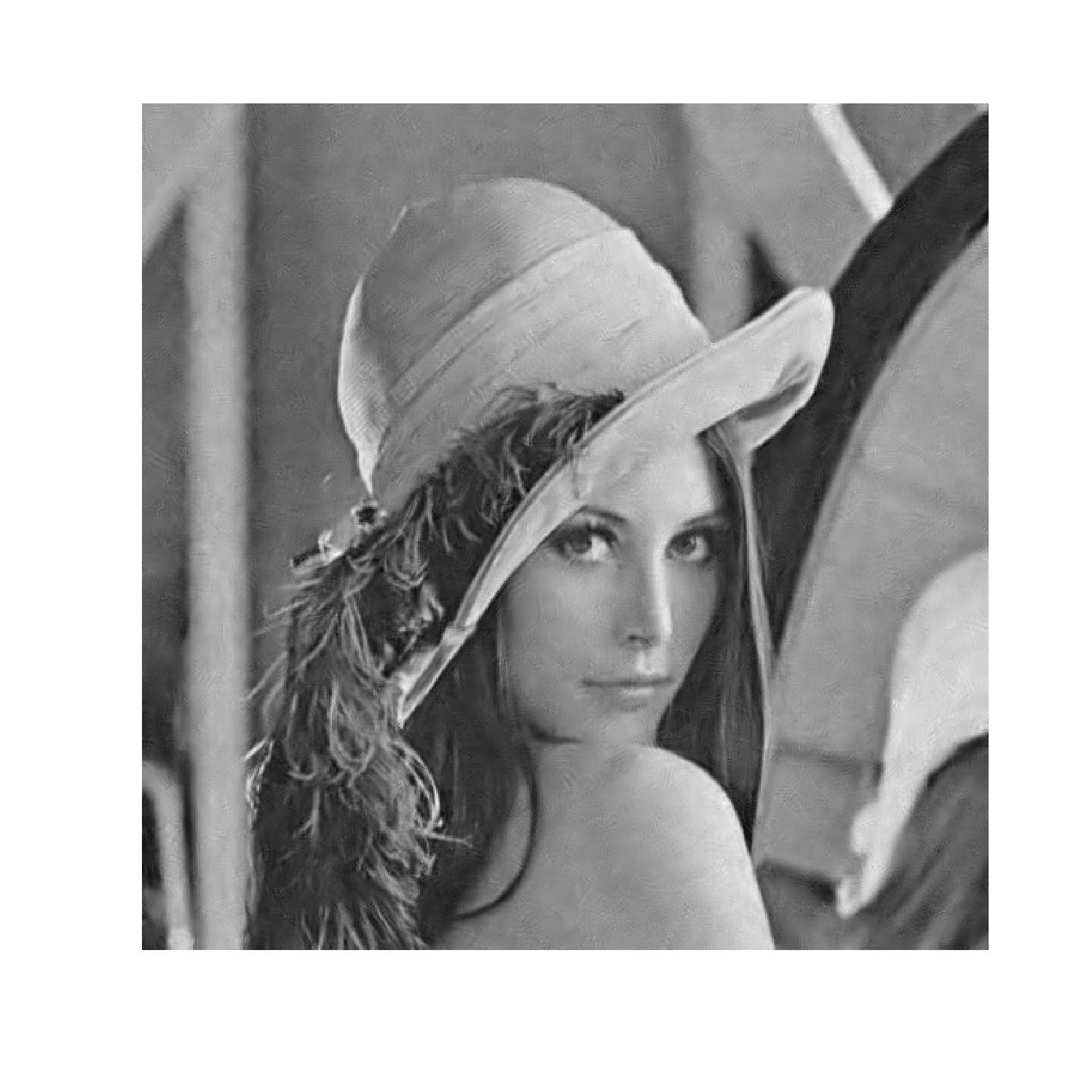}}
\subfigure[{\scriptsize QMPI(32.00dB/0.846)}]{\includegraphics[width=0.16\textwidth]{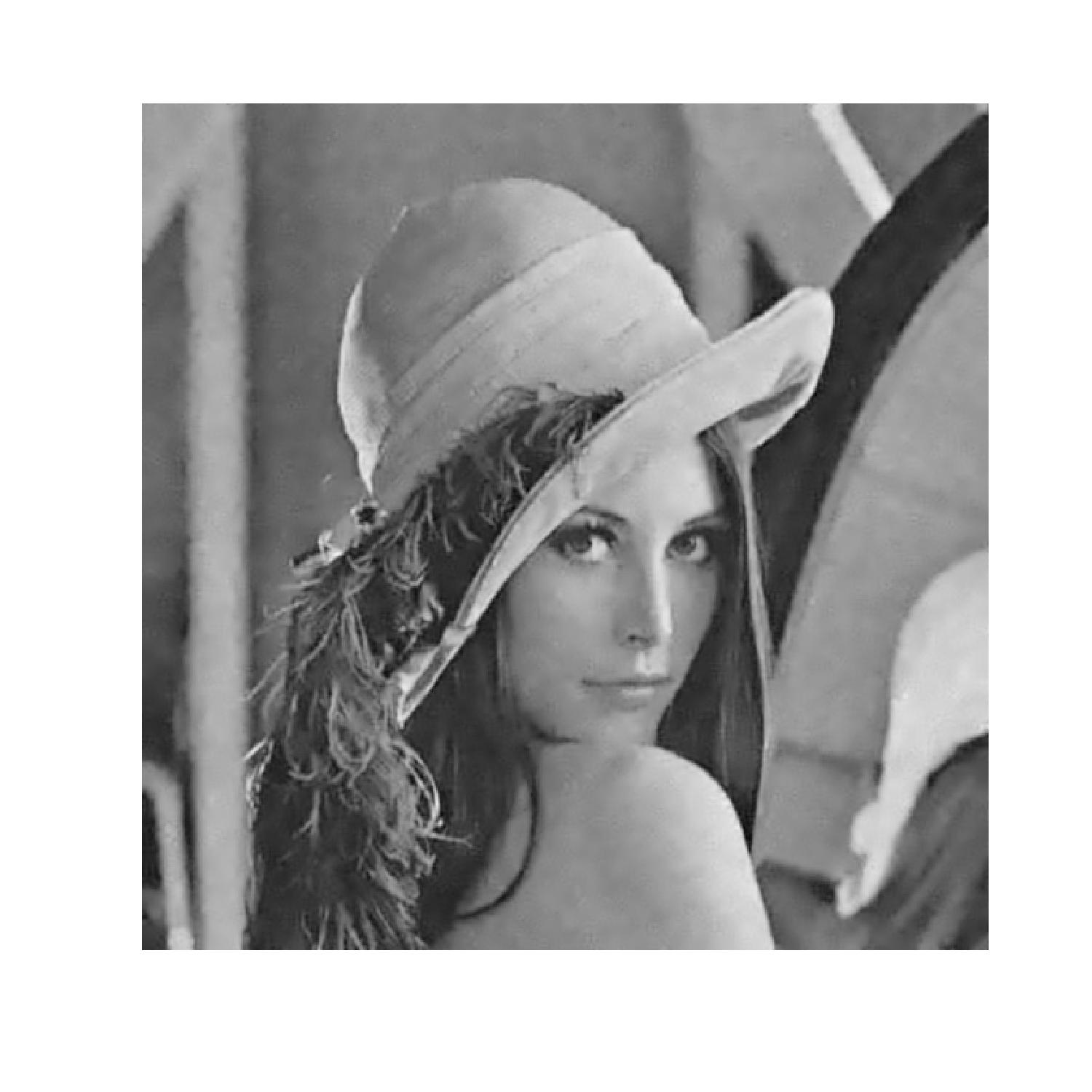}}

\vspace{-3.4mm}
\caption{{\scriptsize Lena image corrupted with 16 dB AWGN. The (PSNR/SSIM) values are noted for all methods. $d = 22$, $p = 0.051$, $\hbar ^2/2m = 1.58$ were used in the proposed method.}}
\label{fig:reslena}
\vspace{-6.3mm}
\end{figure*}

\begin{figure*}[h!]
\centering

\subfigure[{\scriptsize Clean image}]{\includegraphics[width=0.16\textwidth]{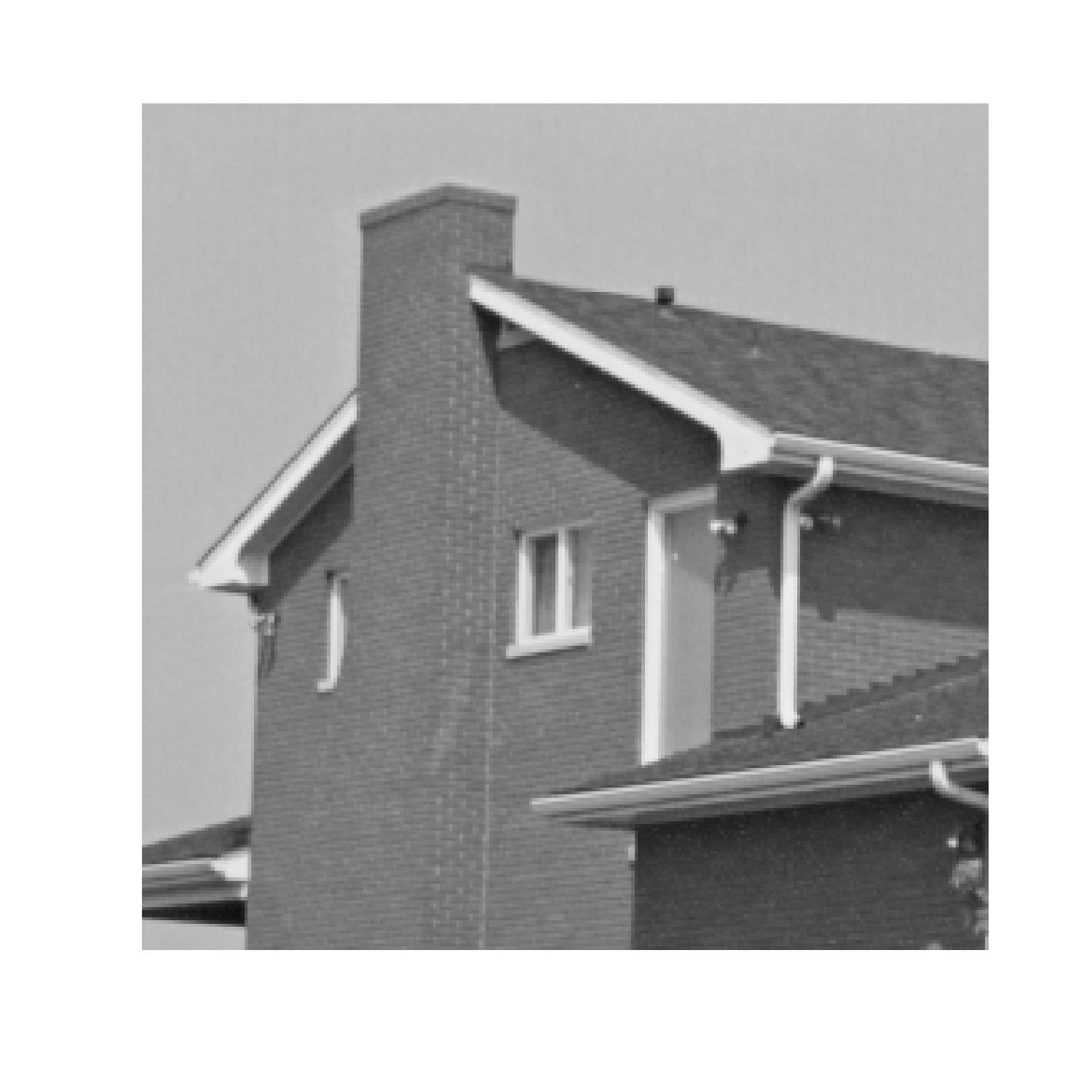}}
\subfigure[{\scriptsize Noisy image(8dB)}]{\includegraphics[width=0.16\textwidth]{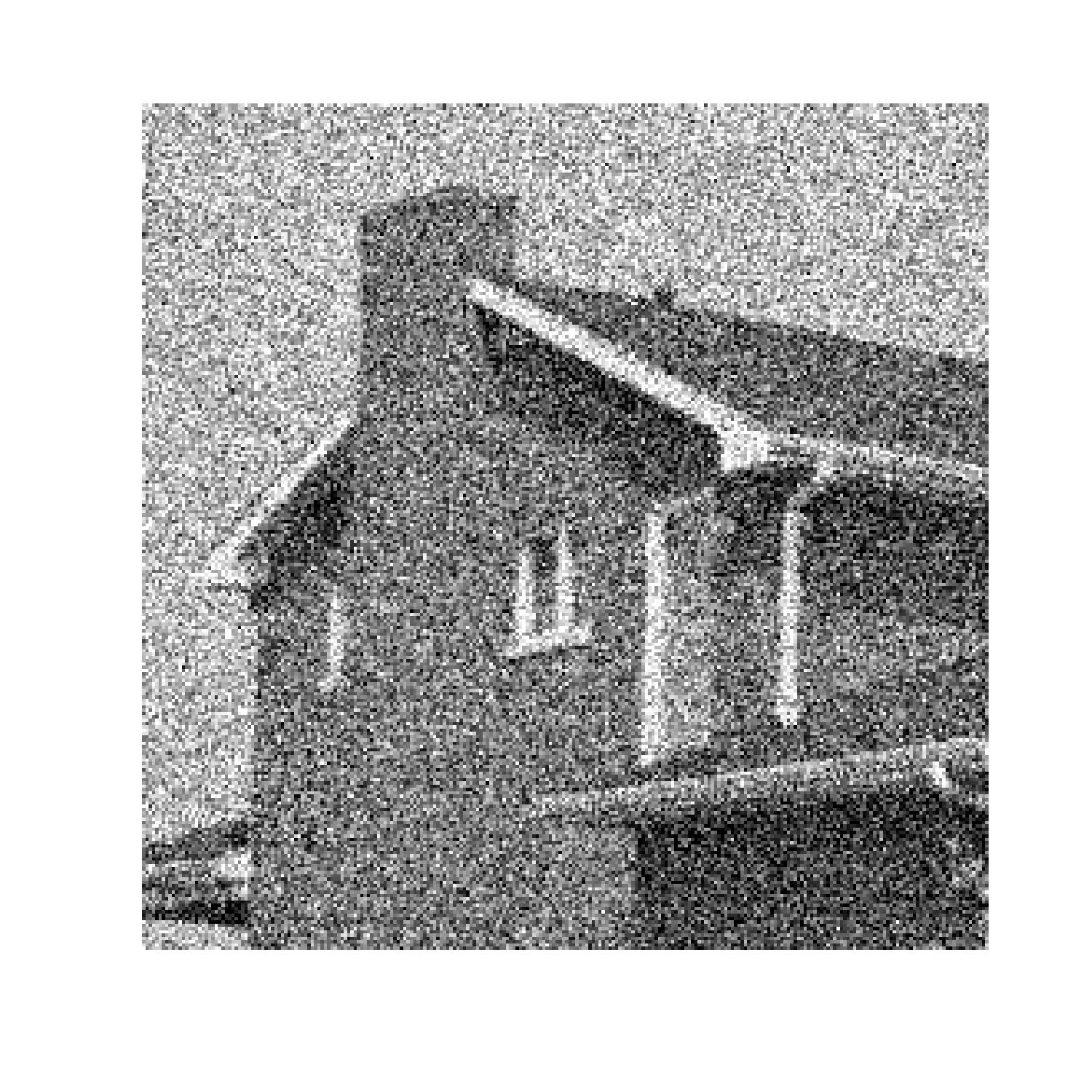}}
\subfigure[{\scriptsize PND(27.35dB/0.751)}]{\includegraphics[width=0.16\textwidth]{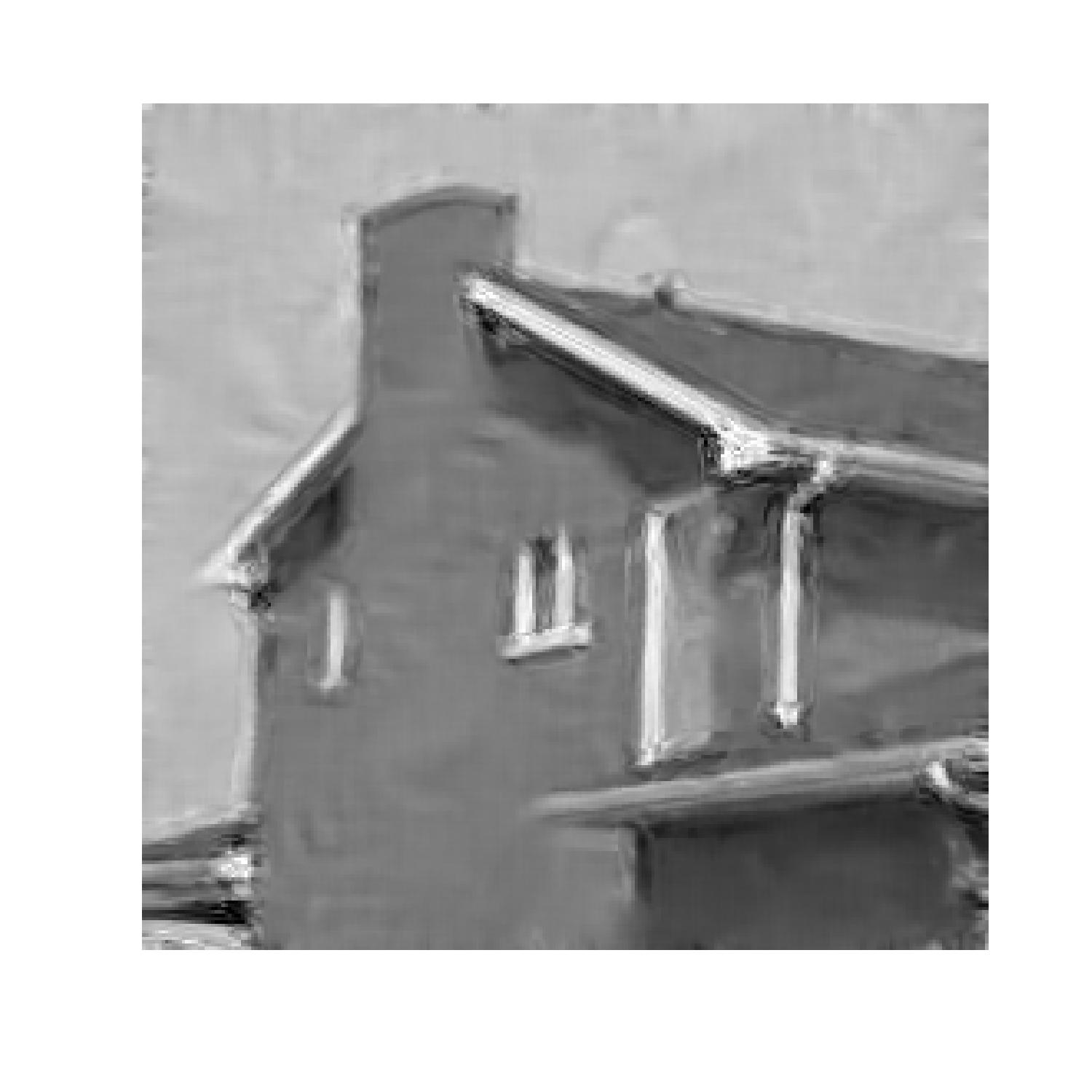}}
\subfigure[{\scriptsize PGPCA(27.13dB/0.687)}]{\includegraphics[width=0.16\textwidth]{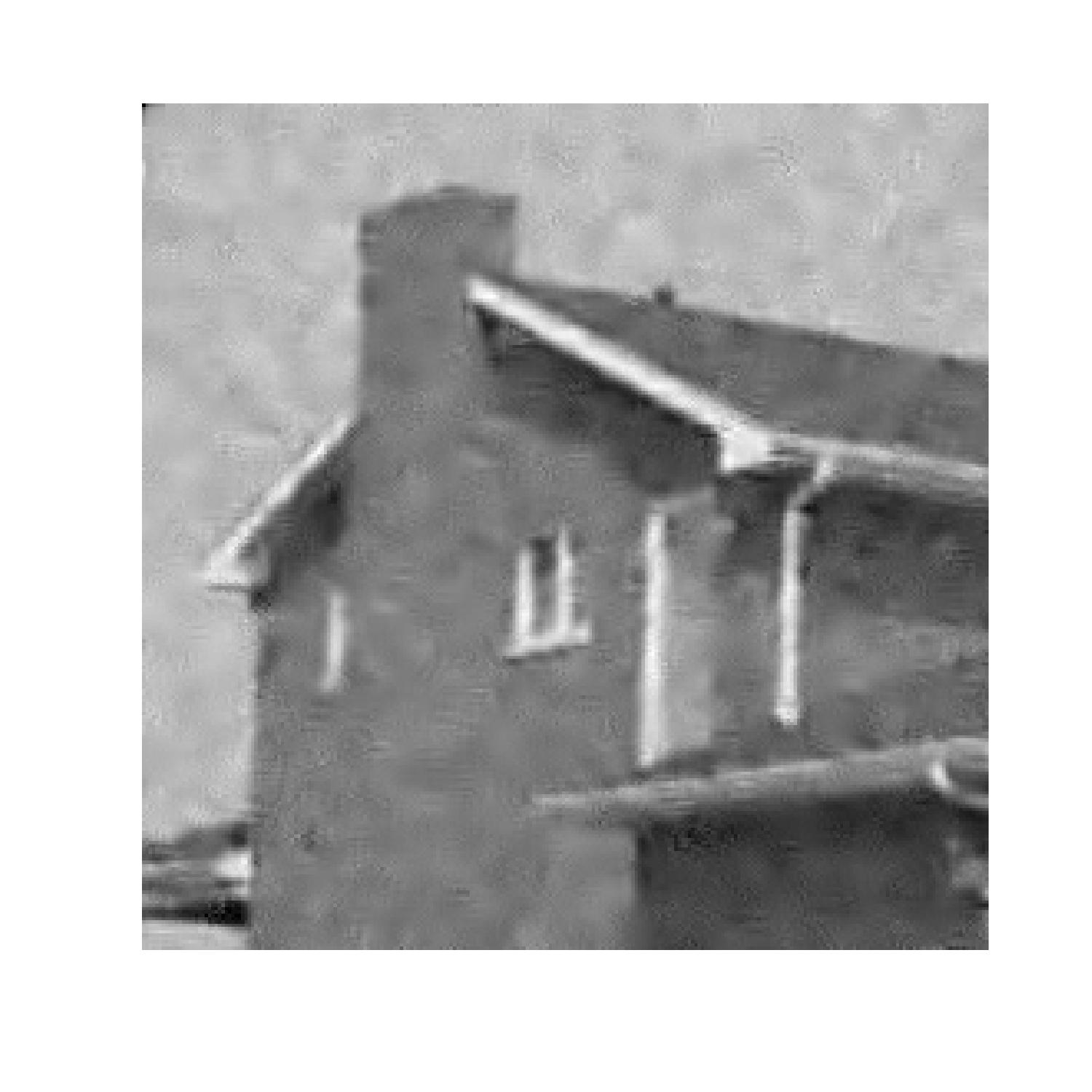}}
\subfigure[{\scriptsize PLPCA(27.04dB/0.648)}]{\includegraphics[width=0.16\textwidth]{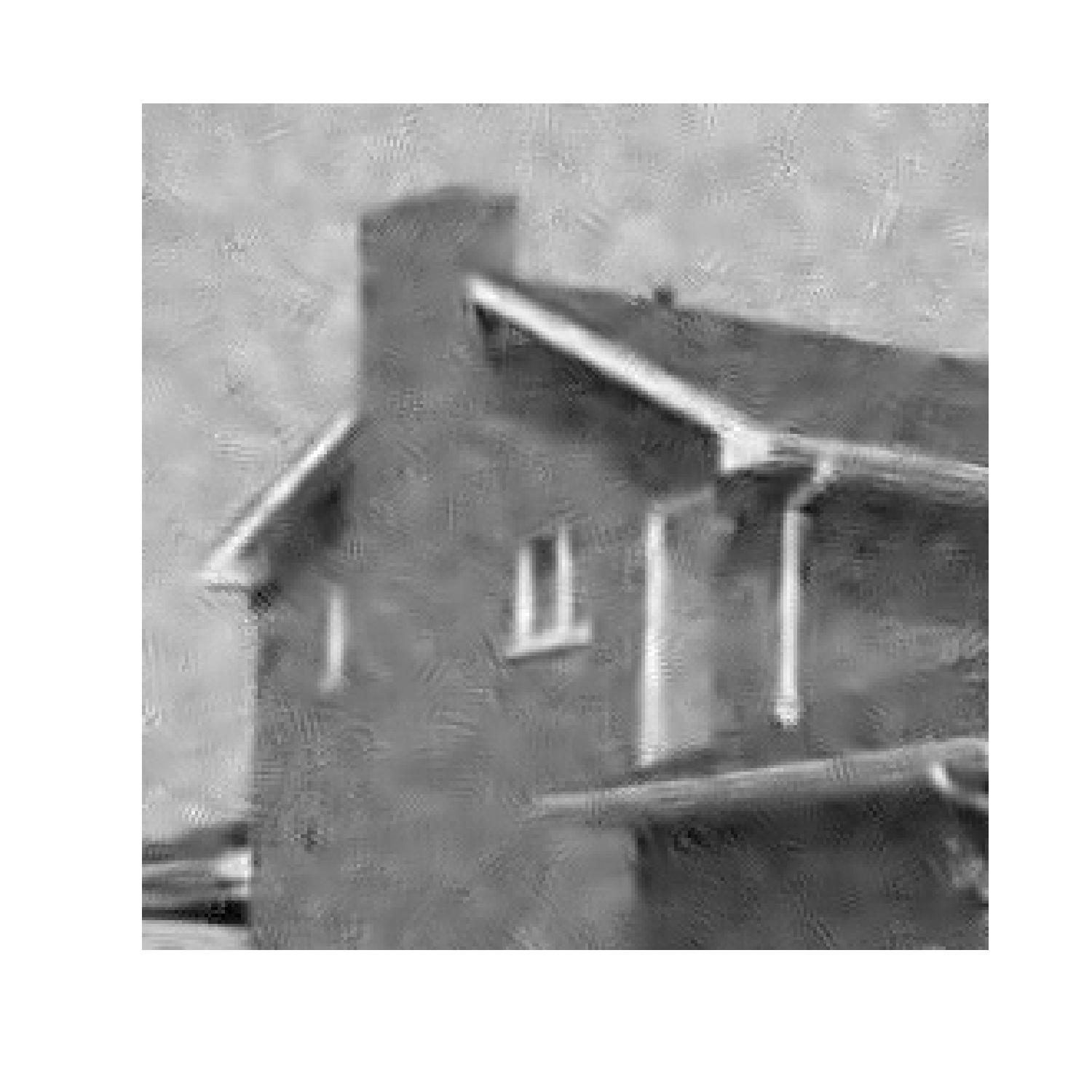}}
\subfigure[{\scriptsize QMPI(27.46dB/0.752)}]{\includegraphics[width=0.16\textwidth]{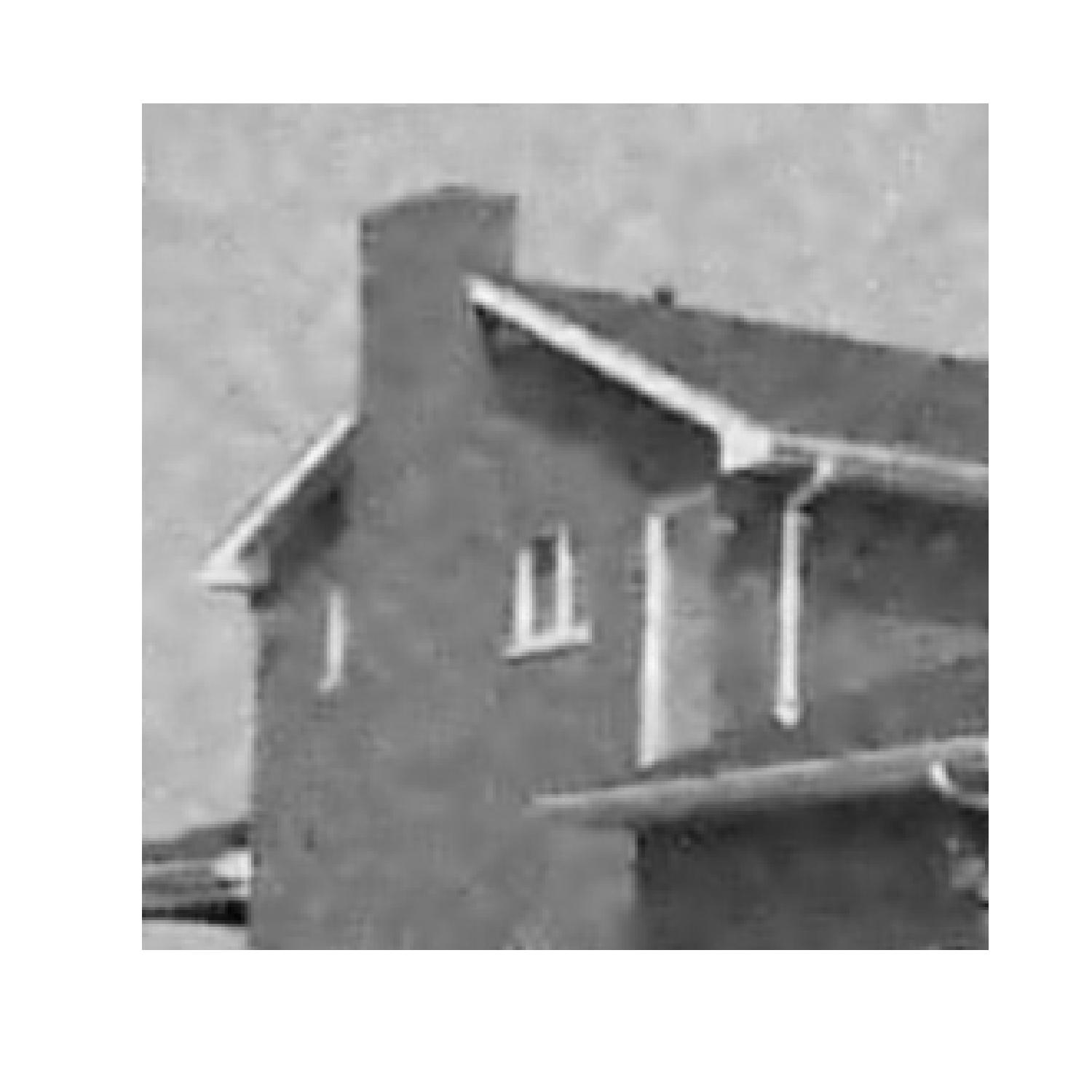}}

\vspace{-3.4mm}
\caption{{\scriptsize House image corrupted with 8 dB AWGN. The (PSNR/SSIM) values are noted for all methods. $d = 11$, $p = 0.085$, $\hbar ^2/2m = 1.53$ were used in the proposed method.}}
\label{fig:reshouse}
\vspace{-6.3mm}
\end{figure*}

\begin{figure*}[h!]
\centering

\subfigure[{\scriptsize Clean image}]{\includegraphics[width=0.16\textwidth]{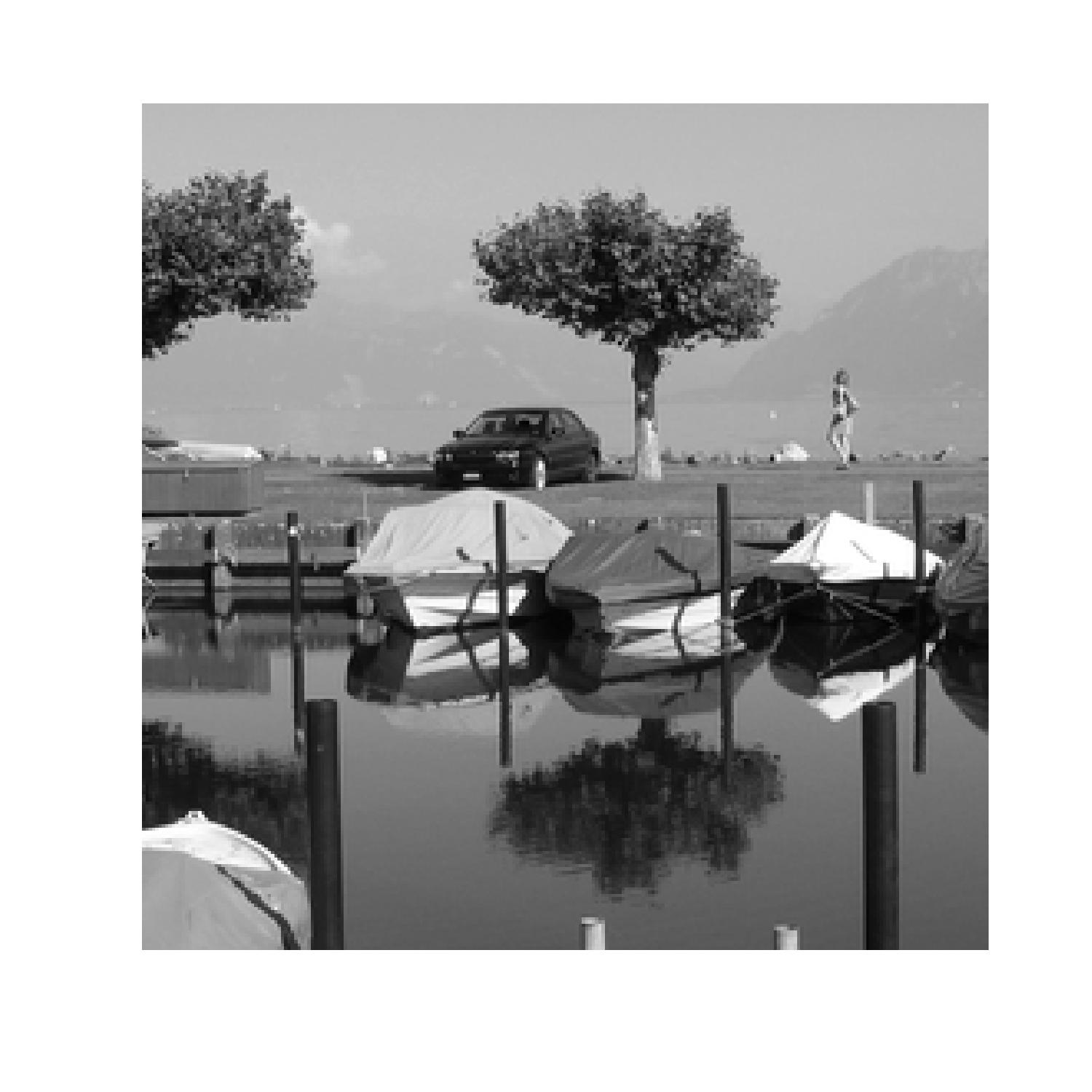}}
\subfigure[{\scriptsize Noisy image(2dB)}]{\includegraphics[width=0.16\textwidth]{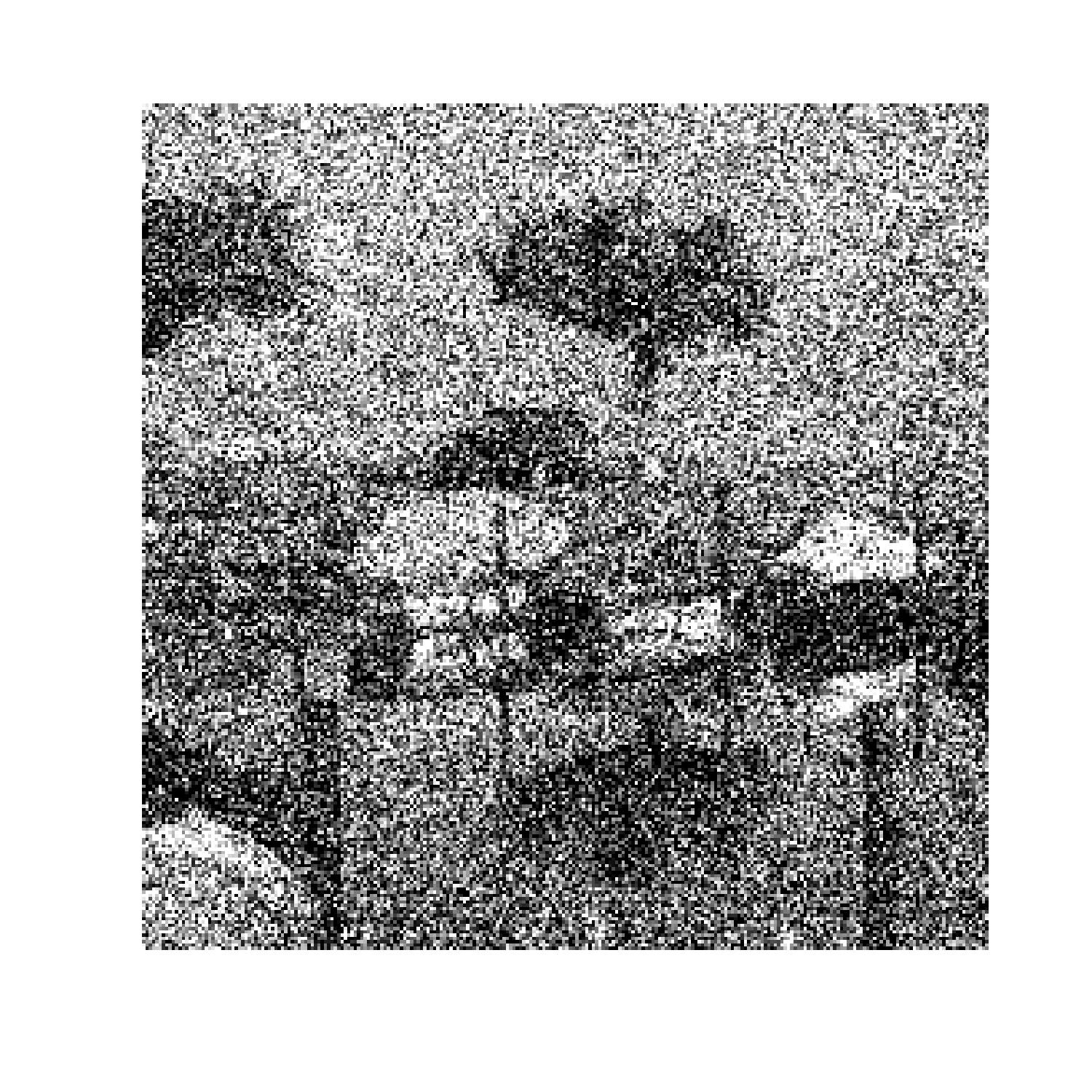}}
\subfigure[{\scriptsize PND(21.48dB/0.608)}]{\includegraphics[width=0.16\textwidth]{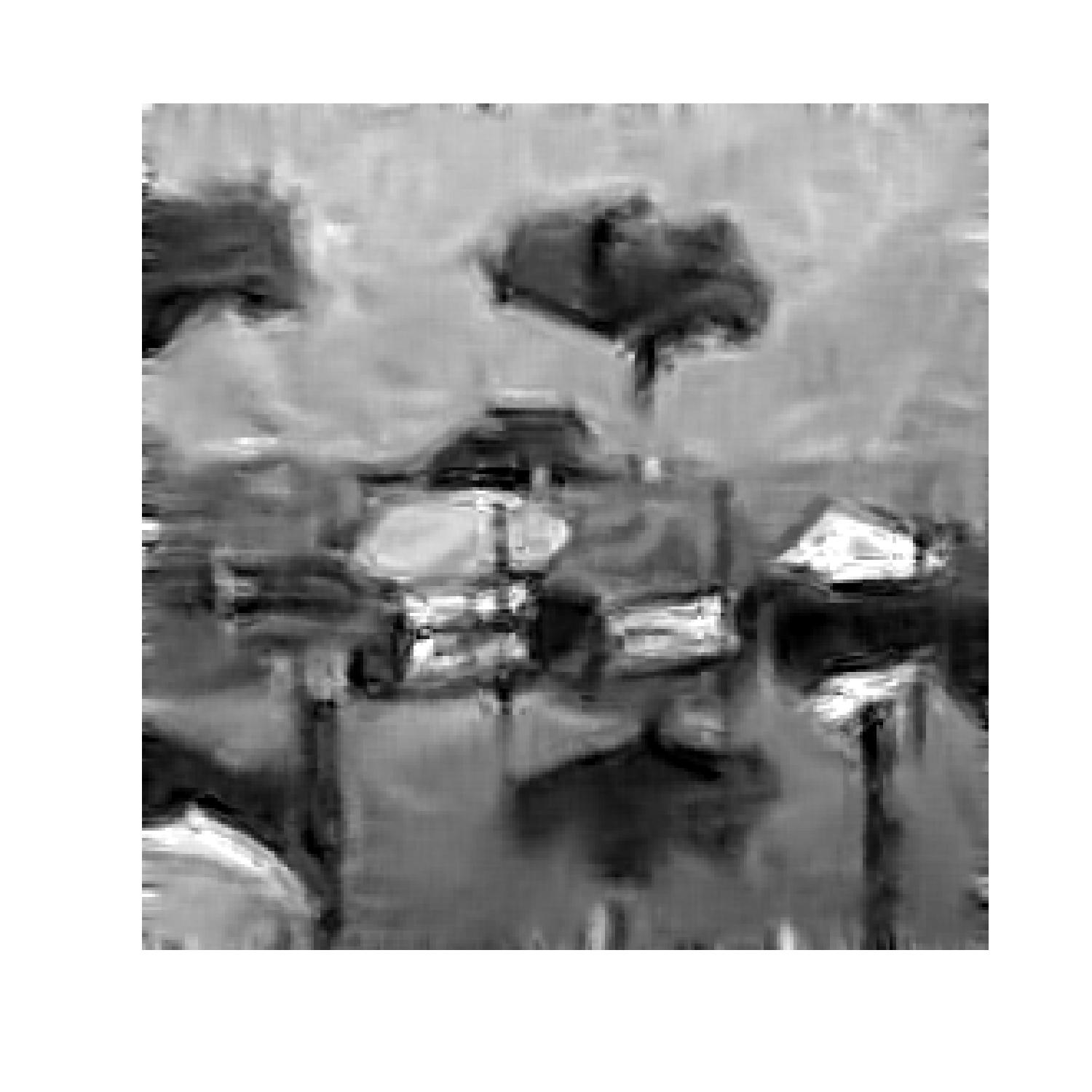}}
\subfigure[{\scriptsize PGPCA(20.97dB/0.460)}]{\includegraphics[width=0.16\textwidth]{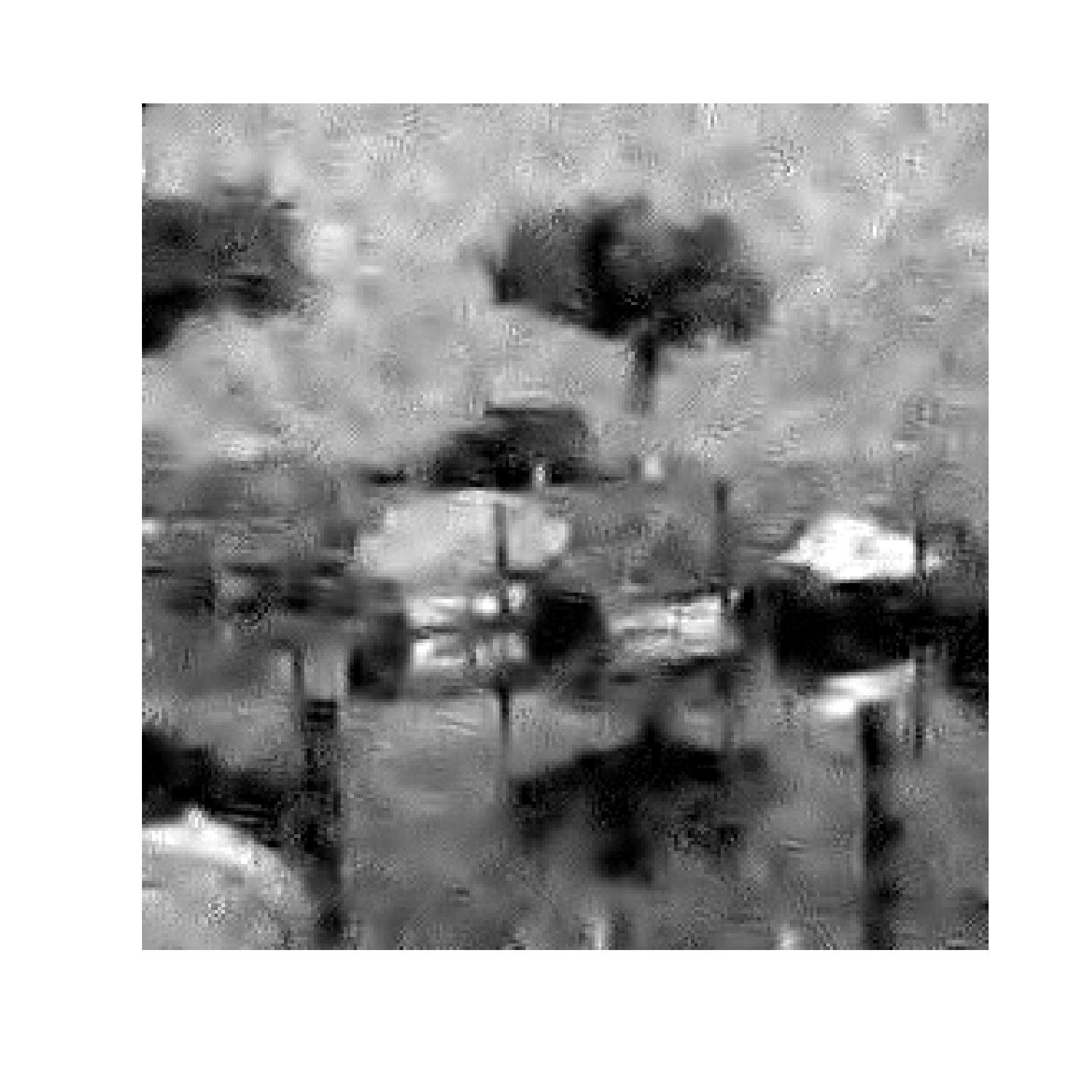}}
\subfigure[{\scriptsize PLPCA(20.57dB/0.406)}]{\includegraphics[width=0.16\textwidth]{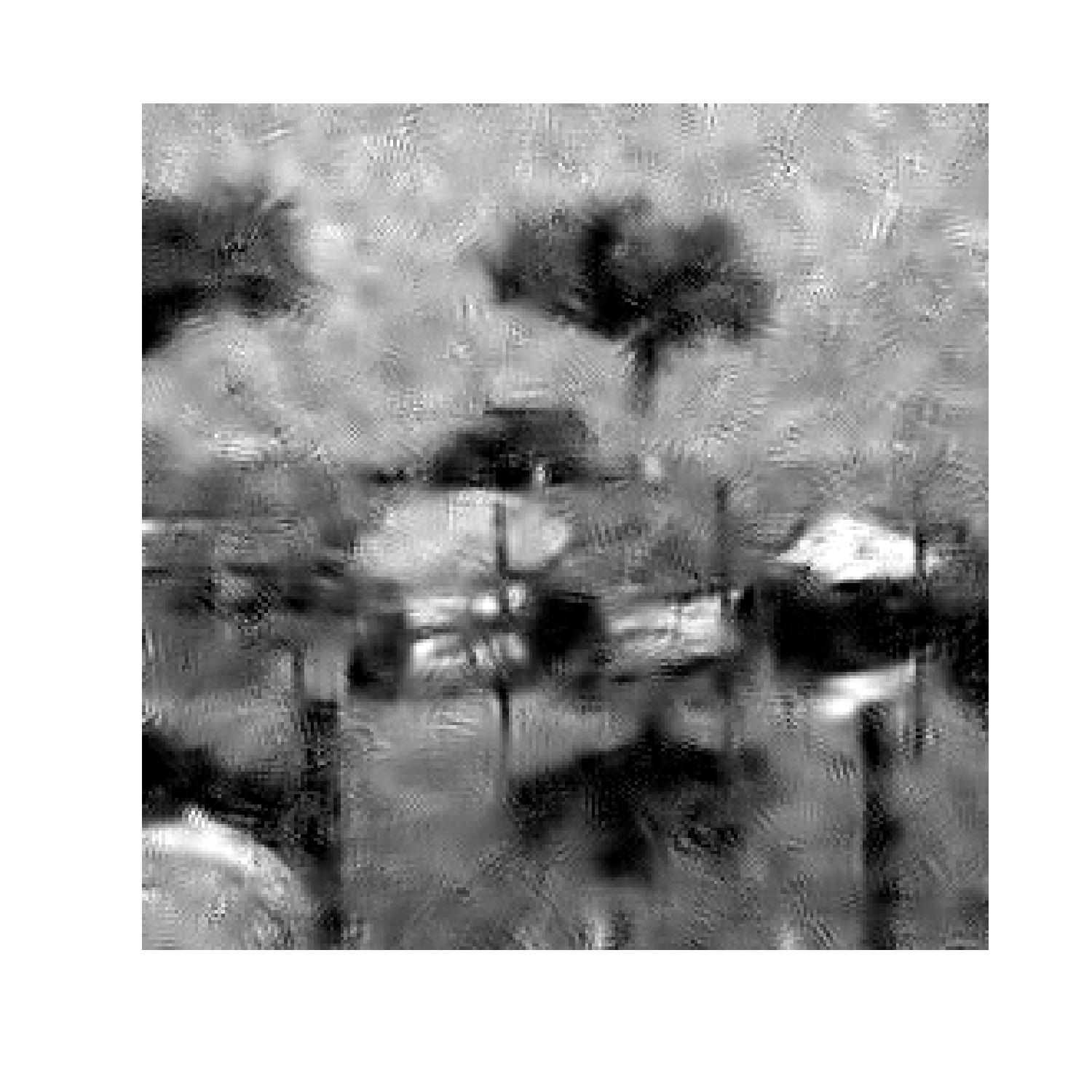}}
\subfigure[{\scriptsize QMPI(21.59dB/0.621)}]{\includegraphics[width=0.16\textwidth]{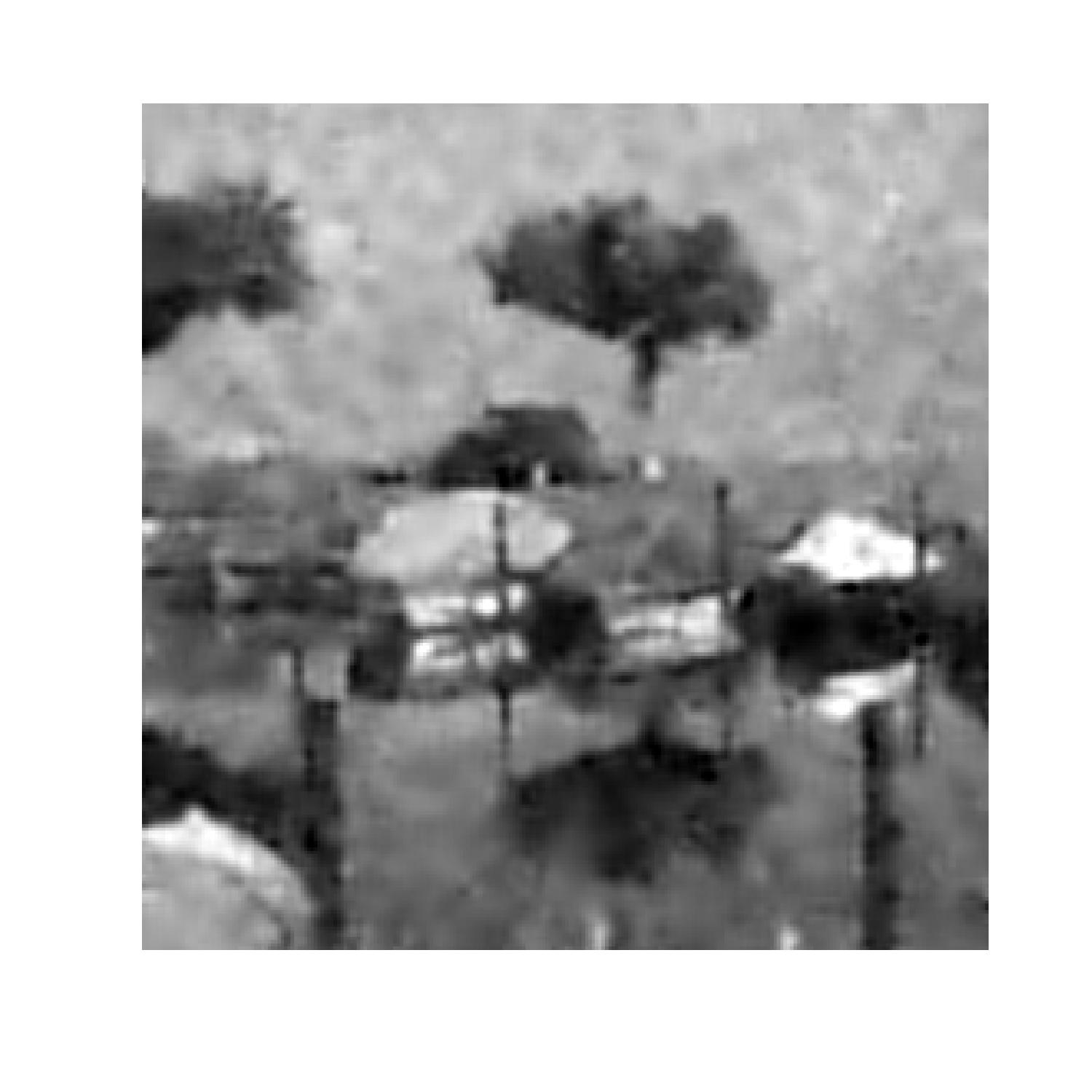}}

\vspace{-3.4mm}
\caption{{\scriptsize Lake image corrupted with 2 dB AWGN. The (PSNR/SSIM) values are noted for all methods. $d = 7$, $p = 0.29$, $\hbar ^2/2m = 2.3$ were used in the proposed method.}}
\label{fig:reslake}
\vspace{-7mm}
\end{figure*}

\begin{table}[h!]
\begin{footnotesize}
\begin{center}
\caption{Quantitative results: PSNR(dB)/SSIM}
\vspace{-3.4mm}

\label{tab:tab_psnr}
\begin{tabular}{c c c c c}
\hline

\multirow{2}{*}{Sample} & \multicolumn{4}{c}{{\scriptsize Methods}}\\
			\cline{2-5}
			 & PND & PGPCA & PLPCA & QMPI \\
\hline\hline
    & \multicolumn{4}{c}{{\scriptsize SNR $\approx$ 22 dB}}\\
	\cline{2-5}

house		& 33.98/0.844 & 35.16/0.883 & \textbf{35.78/0.888} & 35.44/0.884\\
lake			& 30.94/0.865 & 32.87/0.911 & \textbf{33.16/0.913} & \textbf{33.16}/0.912\\
lena 		& 33.88/0.864 & 35.21/0.889 & \textbf{35.52}/0.892 & 35.21/\textbf{0.893} \\

\hline

    & \multicolumn{4}{c}{{\scriptsize SNR $\approx$ 16 dB}}\\
	\cline{2-5}

house		& 31.60/0.814 & 31.73/0.800 & 31.92/0.791 & \textbf{32.15/0.832}\\
lake			& 27.20/0.792 & 28.75/0.808 & \textbf{28.87}/0.803 & 28.85/\textbf{0.821}\\
lena 		& 31.32/0.828 & 31.81/0.815 & 31.89/0.806 & \textbf{32.00/0.846} \\

\hline

    & \multicolumn{4}{c}{{\scriptsize SNR $\approx$ 8 dB}}\\
	\cline{2-5}

house		& 27.35/0.751 & 27.13/0.687 & 27.04/0.648 & \textbf{27.46/0.752}\\
lake			& 23.97/0.703 & \textbf{24.25}/0.652 & 24.07/0.594 & 24.19/\textbf{0.708}\\
lena 		& \textbf{28.07}/0.771 & 27.48/0.705 & 27.28/0.667 & 27.67/\textbf{0.775} \\

\hline

    & \multicolumn{4}{c}{{\scriptsize SNR $\approx$ 2 dB}}\\
	\cline{2-5}

house		& \textbf{24.89/0.686} & 23.14/0.504 & 22.53/0.437 & 23.93/0.682\\
lake			& 21.48/0.608 & 20.97/0.460 & 20.57/0.406 & \textbf{21.59/0.621}\\
lena 		& \textbf{25.16}/0.701 & 23.38/0.517 & 22.75/0.453 & 24.54/\textbf{0.710} \\

\hline
\end{tabular}\end{center}

\end{footnotesize}
\vspace{-9.5mm}
\end{table}


As explained previously, when applied to image denoising, the proposed method borrows the main principle of non local means (NLM) approach. Therefore, comparisons have been carried out with three NLM-based state-of-the-art methods: i) principal component analysis (PCA) for NLM image denoising method called PND in \cite{tasdizen2009principal}, ii) patch-based PCA method for image denoising referred as PGPCA in \cite{deledalle2011image}, and iii) local patch-based PCA method designed for image denoising by collecting patches only from the local neighbourhood labeled as PLPCA in \cite{deledalle2011image}. For all the simulations, the half patch size $P_h$ and half window size $W_h$ were set to 3 and 10 respectively. All the other hyperparameters, for all the methods, have been tuned to provide the best results possible for each experiment. The resulting peak-signal-to-noise-ratio (PSNR) and structure-similarity (SSIM) \cite{Wang2004Image} are regrouped in Table~\ref{tab:tab_psnr}, where best values are highlighted in bold. Figs.~\ref{fig:reslena}-\ref{fig:reslake} illustrate detailed denoising results by these methods for a visual assessment.

PLPCA and PND provide slightly better PSNR values respectively for high and low SNR images. However, the proposed method exhibits better SSIM values within almost all the experiments, justifying its adaptability for high as well as for low SNR images.

\vspace{-4mm}

\section{Conclusions}
\label{sec:conclusion}

This paper introduces an original method for image denoising inspired by the quantum many-body interactive theory. More precisely, an adaptive basis has been constructed using the concept of quantum many-body interaction, which can be used as a filter for denoising the image. The quantum interactions between image patches reflect the local similarities between neighbouring patches of an image. 
We have presented preliminary results showing the interest of this adaptive method for a denoising application in presence of AWGN. These preliminary results show that this new method gives results slightly better than standard 
well-established procedures. A further interest of the method is that it could be applied to any type of noise beyond AWGN without modification. In particular, a perspective of this work could be to apply this method to image-dependent noise models such as Poisson noise \cite{salmon2014poisson}, for which quantum-based methods are well-adapted \cite{dutta2020quantum}.
Another interesting point is that as the denoising is done at the level of individual patches, the computational time is in general much smaller by at least an order of magnitude for large images than in the other quantum-based method of \cite{dutta2020quantum}, and becomes comparable to the one of standard NLM methods.
In addition, another interesting perspective is to extend this idea of quantum interactions for collaborative patch denoising, as originally proposed in \cite{Dabov2007image}. Finally, other image restoration applications (e.g., deconvolution) could also take benefit of quantum interactions through, for instance, plug-and-play algorithms \cite{dutta2020poisson}.

\bibliographystyle{IEEEbib}
\bibliography{bibQMBI}

\end{document}